\documentclass[12pt]{article}
\usepackage{epsfig,amsmath,amsfonts,amssymb}
\usepackage[dvips]{feynmp}
\newcommand{\nc}{\newcommand}
\nc{\rnc}{\renewcommand}
\nc{\N}{{\cal N}}
\nc{\A}{{\cal A}}
\nc{\F}{{\cal F}}
\nc{\Lag}{{\cal L}}
\nc{\G}{{\cal G}}
\nc{\Veff}{{V_{\mbox{\it \!eff}}}}
\nc{\Weff}{{W_{\mbox{\it \!eff}}}}
\nc{\diag}{{\mbox{diag}}}
\nc{\SU}{{\mbox{SU}}}
\nc{\SO}{{\mbox{SO}}}
\nc{\Sp}{{\mbox{Sp}}}
\nc{\U}{{\mbox{U}}}
\nc{\SUSY}{supersymmetry}
\nc{\susy}{supersymmetry}
\nc{\tr}{\mbox{Tr}}
\nc{\Det}{\mbox{Det}}
\nc{\dslash}{\slash\!\!\!}
\nc{\im}{{\mbox{Im}}}

\begin{document}
\rightline{\vbox{\baselineskip12pt\hbox{hep-th/0407084}\hbox{12 Jul 2004}}}

\vskip 1cm

\begin{center}
{\Large \bf Effective Potentials and the Vacuum Structure of Quantum Field Theories}
\vskip 1cm
{\large Kristian D. Kennaway} \\
\vskip .4cm
{\it Department of Physics and Astronomy}\\
{\it University of Southern California}\\
{\it Los Angeles, CA 90089-0484, USA}
\end{center}

\begin{abstract}
I review some older work on the effective potentials of quantum field theories, in particular the use of anomalous symmetries to constrain the form of the effective potential, and the background field method for evaluating it perturbatively.  Similar techniques have recently been used to great success in studying the effective superpotentials of supersymmetric gauge theories, and one of my motivations is to present some of the older work on non-supersymmetric theories to a new audience.  The Gross-Neveu model exhibits the essential features of the techniques.   In particular, we see how rewriting the Lagrangian in terms of an appropriate composite background field and performing a perturbative loop expansion gives non-perturbative information about the vacuum of the theory (the fermion condensate).  The effective potential for QED in a constant electromagnetic background field strength is derived, and compared to the analogous calculation in non-Abelian Yang-Mills theory.  The Yang-Mills effective potential shows that the ``perturbative'' vacuum of Yang-Mills theory is unstable, and the true vacuum has a non-trivial gauge field background.  Finally, I describe how some of the limitations seen in the non-supersymmetric theories are removed by supersymmetry, which allows for exact computation of the effective superpotential in many cases.
\end{abstract}

\section{Introduction}
\label{sec:intro}

Work emerging from string theory over the last few years has led to the computation of the exact low-energy effective superpotential for gauge theories with $\N=1$ \susy\ and matter in various representations.  This generalizes results of Veneziano and Yankielowicz \cite{Veneziano:1982ah}, and Veneziano, Yankielowicz and Taylor \cite{Taylor:1983bp} from the mid 1980s on the effective superpotential in pure $\N=1$ Yang-Mills and $\N=1$ QCD with fundamental matter; in fact, as I shall review, the recent work on $\N=1$ theories is an extension of ideas that have been used for nearly 70 years to study the vacuum structure of quantum field theories in external fields\footnote{In 1936, Euler and Heisenberg computed the 1-loop effective potential of QED in a constant electromagnetic field \cite{Heisenberg:1936qt}.}.

In this paper I review techniques for computing effective potentials in non-supersymmetric field theories, emphasizing the continuity of these techniques with recent work, their limitations in non-supersymmetric theories and how supersymmetry removes some of these limitations.  

When a quantum field theory possesses continuous symmetries, the form of the effective potential (the non-derivative terms in the effective Lagrangian) is constrained by the corresponding (anomalous) Ward identities, which give rise to partial differential equations that must be satisfied by the quantum corrected effective potential.  For example, as we will discuss in section \ref{sec:anom.sym} the differential equation associated to the anomalously broken scaling symmetry is the Callan-Symanzik equation.

The {\it background field method} can be used to derive the one-loop effective action from the path integral of the theory; in theories with non-trivial vacua, such as asymptotically free theories, this gives an approximation to the vacuum state.  Evaluating the 1-loop effective action is equivalent to the summation of an infinite class of Feynman diagrams where one includes the couplings of a set of fluctuating fields to a classical background field, but ignore the self-interactions of the fluctuating fields.

We begin by studying the Gross-Neveu model, a two-dimensional theory of chiral fermions which is asymptotically free.  This model exhibits several of the features of more interesting four-dimensional theories such as Yang-Mills theory and QCD, including asymptotic freedom and spontaneous chiral symmetry breaking.  We will solve for the 1-loop effective potential of this model, as a warm-up exercise for studying four-dimensional gauge theories.

Due to the Landau pole (divergence of the perturbative gauge coupling at low energies), the one-loop approximation to the Yang-Mills effective potential cannot be extrapolated to the vacuum of the theory, but it gives a qualitative picture of some of the features of the vacuum.  When the theory has supersymmetry, the constraints on the effective (super)potential become much more powerful, and the one-loop perturbative gauge theory computations can be extrapolated all the way to low energies to obtain exact, non-perturbative information about the vacuum.

Most of the material reviewed here is well-studied in the literature and dates to the 1970s and 1980s; one of my motivations in writing this review is to present it to an audience who may not be familiar with this older work.  I apologize in advance for errors and omissions resulting from my ignorance of the literature: in particular I have been unable to review much of the subsequent work on non-supersymmetric gauge theories. Conversely, I hope my overview of recent work on supersymmetric theories provides an orientation and introduction to these recent results.

\section{A toy model: the Gross-Neveu model}
\label{sec:gn}

The Gross-Neveu model \cite{Gross:1974jv} is a simple model that exhibits spontaneous symmetry breaking through a quantum-mechanical symmetry-violation.  It is a two-dimensional, asymptotically-free theory of $N$ massless interacting fermions, with Lagrangian:

\begin{equation}
\label{eq:lag.gn}
{\cal L}_{GN} = \overline \psi_i \imath \dslash{\partial} \psi_i + \frac{g^2}{2} (\overline \psi_i \psi_i)^2
\end{equation}
The classical Lagrangian has a discrete chiral symmetry

\begin{equation}
\label{eq:gn.chisym}
\psi_i \rightarrow \gamma_5 \psi_i \quad
\overline \psi_i \rightarrow - \overline \psi_i \gamma_5
\end{equation}
By summing the contribution of Feynman diagrams with vanishing external momenta, we will derive the effective potential of the Gross-Neveu model, and find that the chiral symmetry is spontaneously broken in the quantum theory.  This perturbative 1-loop computation provides exact non-perturbative results about the vacuum of the theory at large $N$.

A useful technique for studying the response of quantum field theories to non-trivial field backgrounds is the {\it background field method}.  One splits the external field into a classical, background field, and a fluctuating quantum field, and then evaluates the path integral perturbatively in the fluctuations around the given background.  We will make full use of the background field method when we study non-Abelian Yang-Mills theory in section \ref{sec:ym}.  Because this technique is non-perturbative in the background field, it can be used to probe for phenomena that are invisible in perturbation theory around the usual zero-field background.

\begin{figure}[t]
\begin{center}
\begin{fmffile}{gnfeyn}
\parbox{60mm}{\begin{center}
	\begin{fmfgraph*}(60,60)
		\fmfleft{i}
		\fmfright{o}
		\fmf{scalar,label=$-\imath g^{2}$,label.side=right}{i,o}
	\end{fmfgraph*}\end{center}
}
\parbox{60mm}{\begin{center}
	\begin{fmfgraph*}(60,60)
		\fmfleft{i}
		\fmfright{o}
		\fmf{fermion,label=$\frac{\imath \dslash p}{p^{2}}$,label.side=right}{i,o}
	\end{fmfgraph*}\end{center}
}
\parbox{60mm}{\begin{center}
	\begin{fmfgraph*}(60,60)
		\fmfleft{i}
		\fmfright{o2,o1}
		\fmf{scalar}{i,v}
		\fmf{fermion}{v,o1}
		\fmf{fermion}{o2,v}
		\fmflabel{\quad-1}{v}
	\end{fmfgraph*}\end{center}
}
\end{fmffile}
\end{center}
\caption{The Feynman rules for the Gross-Neveu Lagrangian (\ref{eq:lag.gn2})}
\label{fig:gnfeyn}
\end{figure}
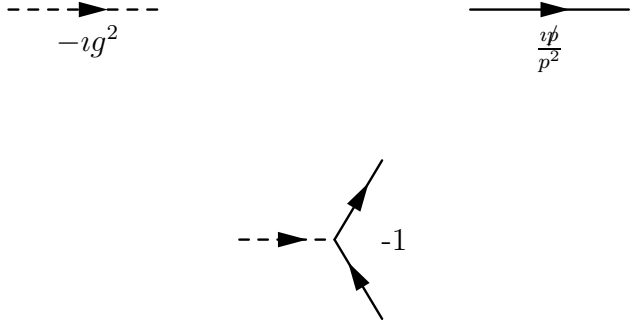

Fermionic (Grassman-valued) fields are not usually considered as classical field theories, for example as possible background fields for a quantum field theory calculation.  However, fermionic quantum fields can pair up and form a composite bosonic field $\sigma \sim \overline \psi  \psi$ which can attain a vacuum expectation value.  The Gross-Neveu Lagrangian can be rewritten as

\begin{equation}
\label{eq:lag.gn2}
\tilde{\cal L} = \overline \psi_i \imath \dslash \partial \psi_i - \frac{1}{2 g^2} \sigma^2 - \sigma \overline\psi_i \psi_i
\end{equation}
which re-expresses it in terms of a coupling to the composite bosonic operator $\sigma$.  This field is treated as a non-dynamical, external background field since it has no kinetic term.  It is easily verified that integrating over this auxiliary $\sigma$ field recovers the original form of the Lagrangian (\ref{eq:lag.gn}).
The Feynman rules for (\ref{eq:lag.gn2}) are shown in figure \ref{fig:gnfeyn}.
 
We will analyze this theory in two ways:  by performing a path integral computation that amounts to summing the Feynman diagrams that can contribute to the effective potential of the theory due to the interaction with the external $\sigma$ field, and by using the anomalously broken scale invariance to constrain the form of quantum corrections to the potential.

\subsection{Path-integral computation of the effective potential}

We can probe the response of the Gross-Neveu model to the formation of a non-zero fermion condensate by introducing an external source $J$ for the field $\sigma$ into the path integral, finding the minimum-energy field configurations in the presence of the source, and then turning off the source.  Define

\begin{equation}
e^{-\imath E[J]} = \int {\cal D} \sigma \prod {\cal D} \psi_{i} \prod {\cal D} \overline \psi_{i}  \exp\left(\imath \left(\Lag(\sigma,\psi_{i},\overline \psi_{i}) + J \sigma\right) \right)
\end{equation}
where $-E[J]$ is the generating functional of connected correlation functions of $\sigma$.  Define the classical field

\begin{equation}
\sigma_{cl}(x) = -\frac{\delta E}{\delta J} = \langle 0|\sigma(x) |0\rangle_{J}
\end{equation}
the vacuum expectation value of $\sigma(x)$ in the presence of the source $J$.  Then the Legendre transform of the energy functional $-E[J]$ defines the effective action $\Gamma(\sigma_{cl})$

\begin{equation}
\Gamma(\sigma_{cl}) = -E[J] - \int d^{4} x~\sigma_{cl}(x) J(x)
\end{equation} 
subject to the constraint
\begin{equation}
\frac{\delta \Gamma(\sigma_{cl})}{\delta \sigma_{cl}(x)} = - J(x)
\end{equation}
Thus, turning off the source $J$ we obtain that the stable configurations for the external field $\sigma_{cl}$ are those for which

\begin{equation}
\frac{\delta \Gamma(\sigma_{cl})}{\delta \sigma_{cl}(x)} = 0
\end{equation}
In the translation-invariant vacuum states of the theory, $\sigma_{cl}(x)$ is constant, and the effective action can be written as

\begin{equation}
\Gamma[\sigma_{cl}] = -(V T) \Veff(\sigma_{cl})
\end{equation}
where $V$ is the 3-dimensional volume, $T$ is the time interval of the integration region, and we defined $\Veff(\sigma_{cl})$ the effective potential for the classical field $\sigma_{cl}$.  The vacua of the theory satisfy

\begin{equation}
\frac{\partial \Veff(\sigma_{cl})}{\partial \sigma_{cl}} = 0
\end{equation}

The effective action is the generating functional of 1-particle irreducible (1PI) correlation functions of the $\sigma$ field.  Therefore in the background of $\sigma_{cl}$

\begin{equation}
\Veff(\sigma_{cl}) = \sum \frac{1}{n\!} \sigma_{cl}^{n} \Gamma_{n}(0,0,\ldots,0)
\end{equation}
where the 1PI diagrams that contribute to $\Gamma_{n}$ carry 0 external momenta on all legs, and each leg comes with a coupling to the background field.  To 1-loop order, the diagrams contributing to the effective potential are shown in figure \ref{fig:gndiags}.  Since they all involve a single fermion loop, we can evaluate the 1PI generating functional to 1-loop order by integrating over the fermions, which appear quadratically in the path integral of the original Lagrangian:

\begin{figure}[t]
\begin{center}
\begin{fmffile}{feyn}
\parbox{20mm}{\begin{center}
	\begin{fmfgraph}(40,40)
		\fmfleft{i}
		\fmfright{o}
		\fmf{dashes}{i,o}
	\end{fmfgraph}\end{center}
} \quad + \quad
\parbox{20mm}{\begin{center}
	\begin{fmfgraph}(40,40)
		\fmfleft{i}
		\fmfright{o}
		\fmf{dashes}{i,v1}
		\fmf{dashes}{v2,o}
		\fmf{fermion,left,tension=0.3}{v1,v2,v1}
	\end{fmfgraph}\end{center}
} \quad + \quad
\def \gnloop#1{%
	\begin{fmfgraph}(40,40)
		\fmfsurroundn{e}{#1}
		\begin{fmffor}{n}{1}{1}{#1}
			\fmf{dashes}{e[n],i[n]}
		\end{fmffor}
		\fmfcyclen{fermion,right=0.2,tension=#1/8}{i}{#1}
	\end{fmfgraph}}
\parbox{20mm}{\begin{center}
	\gnloop{4}
	\end{center}
} \quad + \quad
\parbox{20mm}{\begin{center}
	\gnloop{6}
	\end{center}
} \quad + \quad \ldots
\end{fmffile}
\end{center}
\caption{Diagrams contributing to the 1-loop effective potential for the background field $\sigma$}
\label{fig:gndiags}
\end{figure}
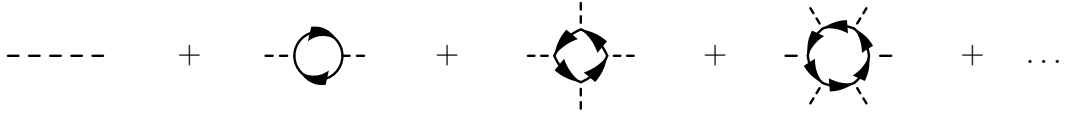

\begin{eqnarray}
Z = \int \prod_i  {\cal D} \psi_i {\cal D} \overline \psi_i {\cal D} \sigma e^{\imath S(\sigma, \psi_i, \overline \psi_i)}
&=& \int \prod_i {\cal D} \psi_i {\cal D} \overline \psi_i  {\cal D} \sigma e^{\imath \int d^2 x \overline \psi_i ( i \dslash{\partial} + \sigma) \psi_i - \frac{\sigma^2}{2 g^2}} \nonumber \\
&=& \int {\cal D} \sigma e^{\imath \int d^2 x \frac{-\sigma^2}{2 g^2}} \Det(\imath \dslash{\partial} + \sigma)^N \nonumber \\
&=& \int {\cal D} \sigma e^{\imath \int d^2 x {\cal L}(\sigma)}
\end{eqnarray} 
with
\begin{equation}
\label{eq:gn.1l}
{\cal L}(\sigma) = - \frac{\sigma^2}{2 g^2} + \imath N \log \Det(\imath \dslash \partial + \sigma)
\end{equation}

Using the relation $\log \Det A = \tr \log A$, we can extract the trace (integral) over the continuous position variables in (\ref{eq:gn.1l}).  Performing a Fourier transform and using the two-dimensional gamma matrices $\gamma^0 = \sigma^2$, $\gamma^1 = \imath \sigma^1$, we then evaluate the remaining matrix determinant:

\begin{eqnarray}
\log \Det(\imath \dslash \partial + \sigma) &=& \int \frac{d^2 p}{(2 \pi)^2} \log \det(\dslash p + \sigma) \nonumber \\
&=& \int \frac{d^2 p}{(2 \pi)^2} \log \det\left(\begin{array}{cc} \sigma & -\imath p_0 + \imath p_1 \\
\imath p_0 + \imath p_1 & \sigma \end{array} \right) \nonumber \\
&=& \int \frac{d^2 p}{(2 \pi)^2} \log (\sigma^2-p^2)
\end{eqnarray}
Therefore

\begin{equation}
\label{eq:gn1loop}
\Lag(\sigma) = - \frac{\sigma^2}{2 g^2}  + \imath N \int \frac{d^2 p}{(2 \pi)^2} \log( \sigma^2 - p^2)
\end{equation}
The 1-loop 1PI correlation functions of (\ref{eq:lag.gn2}) may be derived from ${\cal L}(\sigma)$, and to this order we can identify the Lagrangian $\Lag(\sigma)$ with the effective Lagrangian associated to the effective action $\Gamma(\sigma)=\int d^{d}x~\Lag_{\mbox{\it \!eff}}(\sigma)$, or in other words

\begin{equation}
\Veff(\sigma_{cl}) = - {\cal L}(\sigma_{cl})
\end{equation}
We can recover the diagram sum explicitly by writing
\begin{eqnarray}
\log (\sigma^{2} - p^{2}) &=& \log(-p^{2}(1-\frac{(\imath \sigma)^{2}}{p^{2}})) \nonumber \\
&=& \log(1-\frac{(\imath \sigma)^{2}}{p^{2}}) + \log(-p^{2})\nonumber \\
&\sim& \log(1-\frac{(\imath \sigma)^{2}}{p^{2}}) \nonumber \\
&=& - \sum_{n=1}^{\infty} \frac{1}{n} (\frac{\imath \sigma}{p})^{2n} \nonumber \\
&=& - \tr \sum_{n=1}^{\infty} \frac{1}{n} (- (\imath \sigma) \frac{\imath \dslash p}{p^{2}})^{2n}
\end{eqnarray}
where in the third line we dropped the second term since it just gives rise to an infinite constant upon Wick rotation and integration over $p$.  Comparing to the Feynman rules in figure \ref{fig:gnfeyn}, each term in the series corresponds to a 1-loop diagram of the form shown in figure \ref{fig:gndiags}; therefore, integrating over the fermions to quadratic order is equivalent to computing the 1-loop diagram sum to all orders.  

Returning to the 1-loop effective Lagrangian, the integral (\ref{eq:gn1loop}) is divergent and needs to be regularized.  Rotating the integration contour to continue to Euclidean momenta and using dimensional regularization we obtain

\begin{eqnarray}
\Lag(\sigma) &=& -\frac{\sigma^2}{2 g^2}  + \imath N \int \frac{d^2 p}{(2 \pi)^2} \log( \sigma^2 - p^2)\nonumber \\
&=& -\frac{\sigma^2}{2 g^2} + N \mu^{d-2} \int \frac{d^d p_E}{(2 \pi)^d} (-1)^{-\alpha}\frac{\partial}{\partial \alpha} \left.\left( \frac{1}{p_E^2 + \sigma^2} \right)^{-\alpha}\right|_{\alpha=0} \nonumber \\
&=& -\frac{\sigma^2}{2 g^2} + N \mu^{d-2} \frac{\partial}{\partial \alpha} \left( \frac{(-1)^{-\alpha} \imath}{(4 \pi)^{\frac{d}{2}}} \frac{\Gamma(-\alpha-\frac{d}{2})}{\Gamma(-\alpha)} \left.\left( \frac{-1}{\Delta} \right)^{-\alpha-\frac{d}{2}}\right)\right|_{\alpha=0}
\end{eqnarray}
where $\Delta = \sigma^2$ and $\mu$ is an arbitrary scale introduced to preserve the  uniform scaling of the Lagrangian.  Using the expansion of $\Gamma(x)$ near its poles, $\Gamma(x) \sim \frac{(-1)^n}{n! (x+n)} - \gamma + 1 + \ldots + \frac{1}{n} + {\cal O}(x+n)$ and $\Gamma(x+1) = x \Gamma(x)$
 we expand $\Gamma(-\alpha-\frac{d}{2})$ and write the singular terms in the form suitable for the modified minimal subtraction scheme (adapted to 2 dimensions):
 
\begin{eqnarray}
\frac{\Gamma(1-\frac{d}{2})}{(4 \pi)^{d/2}} \left( \frac{-1}{\Delta} \right)^{1-d/2} &=&\frac{1}{4 \pi} \left( \frac{1}{\epsilon} - \gamma + \log 4 \pi - \log \Delta  + {\cal O}(\epsilon) \right) \nonumber \\
&\longmapsto& -\frac{1}{4 \pi} \log(\frac{\Delta}{\mu^2})
\end{eqnarray}

\begin{figure}[t]
\begin{center}
\epsfig{file=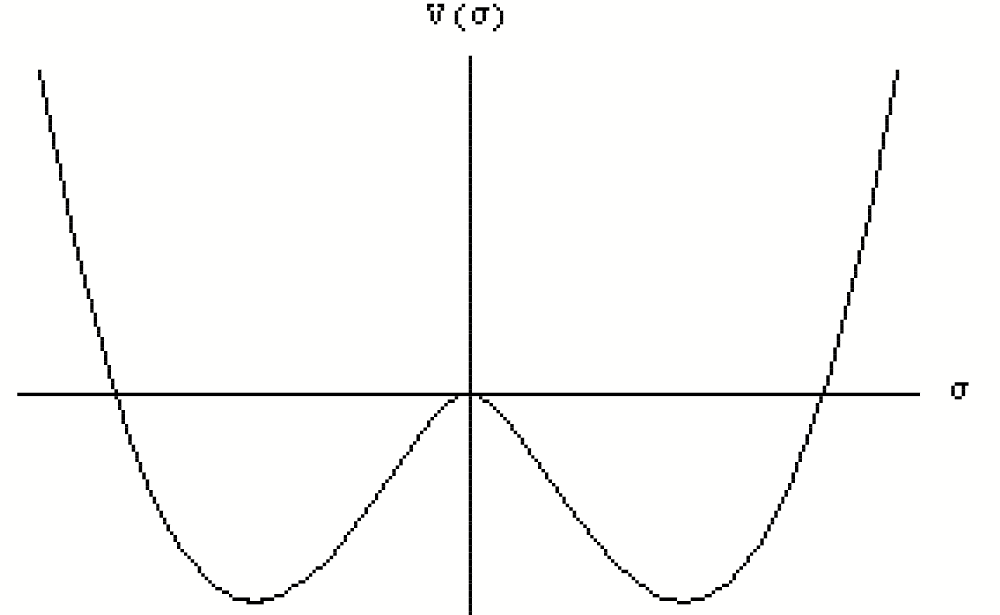}
\parbox{5.5in}{
\caption{1-loop effective potential of the Gross-Neveu model.\label{fig:gn.pot}}}
\end{center}
\end{figure}

We obtain for the effective potential
 
\begin{eqnarray}
\label{eq:gnpot}
\Veff(\sigma_{cl}) &=& \frac{\sigma_{cl}^2}{2 g^2} + \frac{N}{4 \pi} \sigma_{cl}^2 \left( \log \frac{\sigma_{cl}^2}{\mu^2} - 1 \right) \nonumber \\
 &=& \frac{N \sigma_{cl}^2}{4 \pi} \left( \log \frac{\sigma_{cl}^2}{\Lambda^2}  - 1 \right)
\end{eqnarray}
where in the second line we defined the dynamical scale $\Lambda^2 = \mu^2 \exp(\frac{-2 \pi}{N g^2})$.  The potential (\ref{eq:gnpot}) is of Coleman-Weinberg type \cite{Coleman:1973jx} and has the form shown in figure \ref{fig:gn.pot}.
 
Extremizing (\ref{eq:gnpot}), we find that what was the classical minimum $\langle \sigma_{cl} \rangle = 0$ is now a local maximum, and there are degenerate global minima at $\langle \sigma_{cl} \rangle = \pm \Lambda$.  Thus, the original ``perturbative'' vacuum can minimize its energy by spontaneously generating a background of paired fermions,

\begin{equation}
\langle \overline \psi \psi \rangle = \frac{1}{g^2} \langle \sigma \rangle = \pm \frac{\mu}{g^2} \exp(\frac{-\pi}{N g^2})
\end{equation}
and since this fermion bilinear does not respect the chiral symmetry (\ref{eq:gn.chisym}), the Gross-Neveu model exhibits spontaneous chiral symmetry breaking.  

Higher loop corrections to the effective potential necessarily involve $\sigma$ propagators and are therefore suppressed by powers of $g$; in fact all higher loop corrections vanish in the 't Hooft limit $N \rightarrow \infty$, $g \rightarrow 0$, $g^{2} N = \mbox{\it const.}$ \cite{Gross:1974jv}.  Therefore, in this limit the 1-loop result is exact.  Unfortunately, for most interesting non-supersymmetric theories (such as Yang-Mills or QCD) the higher-loop corrections do not vanish in this limit, and the infinite diagram series cannot be summed explicitly even at large $N$ \footnote{Although a generating function that enumerates the infinite series of Feynman diagrams is known for QCD \cite{'tHooft:1998mh}}; the complication comes from performing the loop momentum integrals at higher orders.  However we can obtain partial results by organizing the diagrams as a loop expansion: in section \ref{sec:ym} we will show how summing the one-loop diagrams for Yang-Mills theory in a covariantly constant background field strength gives a (not particularly good) approximation to the vacuum state of Yang-Mills theory.

However, simplifications even more powerful than those of the Gross-Neveu model were observed recently in certain four-dimensional $\N=1$ theories, where supersymmetry provides additional constraints on the effective potential that allows us to sum the diagram expansion to all orders.  We will come back to this in section \ref{sec:susy.matter}.

The value of the fermion condensate $\langle \overline \psi \psi \rangle = \pm \frac{\mu}{g^2} \exp(\frac{- \pi}{N g^2})$ is a non-perturbative quantity, since its Taylor expansion around $g=0$ vanishes to all orders.  Therefore the non-trivial vacuum of the Gross-Neveu model is invisible in the perturbation theory of the original Lagrangian (\ref{eq:lag.gn}), which preserves chiral symmetry to all orders.  It was only by rewriting the Lagrangian by introducing a coupling to the appropriate background field that we could probe for the existence of a chiral symmetry breaking condensate.  We have seen that by introducing an appropriate variable in which to perform a perturbative loop expansion (the composite background field $\sigma$), we can obtain non-perturbative information about the vacuum of the theory, order by order in the perturbative evaluation of a {\it different} Lagrangian.

\subsection{Anomalous symmetries and effective potentials}
\label{sec:anom.sym}

In quantum field theories, continuous symmetries of the classical Lagrangian may sometimes be violated in the quantum theory.  An example of an anomalous symmetry are scale transformations (dilatations) in massless field theories\footnote{Another anomalous symmetry is axial rotations of massless Dirac fermions in gauge theories; the corresponding effective Lagrangian including quantum corrections from the axial anomaly can be obtained by similar techniques, and has been used to study the role of the anomaly in the low-energy dynamics of mesons \cite{DiVecchia:1980ve,Witten:1980sp}.}.  The continuous dilatation symmetry is associated to a current $D_\mu = \Theta^{\mu \nu} x_\nu$, where $\Theta^{\mu \nu}$ is the stress-energy tensor of the theory, defined by

\begin{equation}
\Theta^{\mu \nu} = 2 \frac{\delta}{\delta g_{\mu \nu}(x)} \int d^d x {\cal L}
\end{equation}

Classically the dilatation current is conserved; $\partial_{\mu} D^{\mu} = \Theta^{\mu}_{\mu} = 0$.  However under a change of renormalization scale this symmetry is broken by the running of the coupling constant (see \cite{Peskin:1995ev}), and the one-loop trace anomaly is given by:

\begin{equation}
\partial_\mu D^\mu = \beta(g) \frac{\partial}{\partial g} {\cal L}
\end{equation}
The trace anomaly receives contributions from all orders in perturbation theory, as well as possible non-perturbative corrections, through the beta function.

In a quantum field theory the ``charge'' of fields under a scale transformation (their scaling dimension) may receive quantum corrections as we change the renormalization scale; operators can have anomalous dimensions.  The Callan-Symanzik equation encodes the scaling behavior of the effective potential under a change of renormalization scale (renormalization group invariance):

\begin{equation}
\label{eq:cs}
\left[ d - \sum_i (d_i + \gamma_{{\cal O}_i})  {\cal O}_i \frac{\partial}{\partial {\cal O}_i} +  \beta(g) \frac{\partial}{\partial g} + \mu \frac{\partial}{\partial \mu} \right] \Veff\,({\cal O}_i, g, \mu)  = 0
\end{equation}
where $d$ is the space-time dimension, $d_i$ are the classical scaling dimensions of the fields ${\cal O}_i$ that appear in the effective potential, $\gamma_{{\cal O}_i}$ are their anomalous dimensions, and $\mu$ is the renormalization scale.  This equation imposes that the effective potential must scale with dimension $d$, and reproduce the trace anomaly under a scale transformation.

In order to use the Callan-Symanzik to obtain predictions about the form of $\Veff$, we need to know the $\beta$ function and anomalous dimensions $\gamma$.   These are typically only known through explicit loop calculations, such as the one we did in the previous section.  However, as we will discuss in section \ref{sec:ym.anom}, once we know $\beta$ and $\gamma$ from a particular calculation, we can use the Callan-Symanzik equation to constrain the allowed form of the effective potential for an {\it arbitrary} field background.

We impose
\begin{equation}
\left[2 - (1 + \gamma_{\sigma}) \sigma \frac{\partial}{\partial \sigma} + \beta(g) \frac{\partial}{\partial g} + \mu \frac{\partial}{\partial \mu} \right] \Veff = 0
\end{equation}
and find that $\beta(g)=-\frac{N g^3}{2 \pi}$, $\gamma_\sigma = 0$ \footnote{The field $\sigma$ has vanishing anomalous dimension due to the normalization of the Lagrangian (\ref{eq:lag.gn2}).  A wavefunction renormalization of $\sigma$ cannot be balanced by a coupling-constant renormalization since the coefficient of the $\sigma$ interaction term is fixed to $1$.}. As we noted in the previous section, in the 't Hooft limit these quantities are exact.

\section{Four-dimensional gauge theories}

Before considering non-Abelian Yang-Mills theory, it is instructive to review the calculation of the effective potential for QED in external electromagnetic fields, which shares many technical features with the Yang-Mills case.  These results were first obtained by Euler and Heisenberg in 1936 \cite{Heisenberg:1936qt}, and were cast in a rigorous quantum field theory framework by Schwinger in 1951 \cite{Schwinger:1951nm}.  The presentation here includes elements from \cite{Salam:1975xe,Flyvbjerg:1980qv}.

\subsection{QED}
\label{sec:qed}

The Lagrangian of QED is

\begin{equation}
{\cal L} = -\frac{1}{4} F_{\mu \nu} F^{\mu \nu} + \overline \psi \dslash D \psi + m  \overline \psi \psi
\end{equation}
where the covariant derivative is $D_{\mu} = \partial_\mu + i e A_\mu$.  As in the previous section, the effective action for the gauge field is given to 1-loop order by

\begin{eqnarray}
e^{i \Gamma[A]} &=& \int {\cal D} \overline \psi {\cal D} \psi e^{\imath \int d^4 x {\cal L}} \nonumber \\
&=& \det(\imath \dslash D - m) e^{-\frac{\imath}{4} \int d^4 x  F^2} \nonumber \\
&=& \exp(\imath {{\int d^4 x {{\cal L}_{\mbox{\it \!eff}}}}})
\end{eqnarray}
where we defined the 1-loop effective Lagrangian

\begin{eqnarray}
{\cal L}_{\mbox{\it \!eff}} &=& -\frac{1}{4} F_{\mu \nu} F^{\mu \nu} - \imath \log \det(\imath \dslash \partial - e\dslash A - m) \nonumber \\
\end{eqnarray}

For comparison to Yang-Mills theory in the next section, we henceforth restrict to massless electrons, although the massive case can be easily treated in a similar manner.  To evaluate the fermion determinant $\det(\imath \dslash D)$ it is convenient to evaluate the determinant of $(\imath \dslash D)^2$ and take the square root.  Expanding and using the anticommutation relation $\{ \gamma^\mu, \gamma^\nu\} = 2 g^{\mu \nu}$, we find

\begin{eqnarray}
\label{eq:qed.det}
(\imath \dslash D)^2 = -D^2 - \frac{e}{2} \sigma_{\mu \nu} F^{\mu \nu}
\end{eqnarray}
where $\frac{\imath}{2} [ \gamma^\mu, \gamma^\nu ] = \sigma^{\mu \nu}$ is the generator of Lorentz transformations on the spin-$\frac{1}{2}$ representation.  Therefore

\begin{equation}
\log \det( \imath \dslash D) = \frac{1}{2}  \log \det(-D^2 - \frac{e}{2} \sigma_{\mu \nu} F^{\mu \nu})
\end{equation}

As we discussed in the previous section, the determinant corresponds to summing up the infinite series of 1-loop Feynman diagrams of the theory, where the electron runs in the loop, and we consider arbitrary insertions of the background gauge field.  The one-loop effective Lagrangian for massless QED is therefore

\begin{equation}
{\cal L}_{\mbox{\it \!eff}} = -\frac{1}{4} F_{\mu \nu} F^{\mu \nu} - \frac{\imath}{2} \tr \log((p_\mu - A_\mu)^2 - \frac{e}{2} \sigma_{\mu \nu} F^{\mu \nu})
\end{equation}
This Lagrangian exhibits the anomalous magnetic moment interaction $\frac{e}{2} \sigma_{\mu \nu} F^{\mu \nu}$ of the electron with the background electromagnetic field.  A similar magnetic moment interaction for the charged gluons of Yang-Mills theory will be vital for understanding the vacuum properties of that theory.

In diagonalizing this operator one needs the eigenvalues of the field strengths $F_{\mu \nu}$.  Defining the Lorentz scalar and pseudo-scalars

 \begin{eqnarray}
 \F &=& \frac{1}{4} F_{\mu \nu} F^{\mu \nu} = \frac{1}{2}(B^2-E^2) \nonumber \\
 \G &=& \frac{1}{4} F_{\mu \nu} {\tilde F}^{\mu \nu} = E \cdot B
 \end{eqnarray}
where $\tilde F^{\mu \nu} = \frac{1}{2} \imath \epsilon^{\mu \nu \rho \sigma} F_{\rho \sigma}$ is the dual field-strength tensor.  Using the identities

\begin{eqnarray}
F_{\mu \rho} {\tilde F}^{\rho \nu} &=& -\delta_{\mu}^{\nu} \G \\
{\tilde F}_{\mu \rho} {\tilde F}^{\rho \nu } - F_{\mu \rho} F^{\rho \nu} &=& 2 \delta_{\mu}^{\nu} \F
\end{eqnarray}
the eigenvalues $\lambda$ of $\F_{\mu \nu}$ are found to satisfy

\begin{equation}
\lambda^4 + 2 \F \lambda^2 - \G^2 = 0
\end{equation}
which has solution $\pm \lambda^{(1)}$, $\pm \lambda^{(2)}$, with
\begin{eqnarray}
\label{eq:qed.ev}
\lambda^{(1)} &=& \frac{\imath}{\sqrt{2}} ((\F + \imath \G)^{1/2} + (\F - \imath \G)^{1/2}) \\
\lambda^{(2)} &=& \frac{\imath}{\sqrt{2}} ((\F + \imath \G)^{1/2} - (\F - \imath \G)^{1/2})
\end{eqnarray}
The magnetic moment operator satisfies

\begin{equation}
(\frac{1}{2} \sigma_{\mu \nu} F^{\mu \nu})^2 =2 (\F + \gamma_5 \G)
\end{equation}
therefore using $\gamma_5^2 = -1$ and (\ref{eq:qed.ev}) the eigenvalues are

\begin{equation}
\pm (2 (\F \pm \imath \G))^{1/2}
\end{equation}

In a particular Lorentz frame, a constant magnetic field may be specified by taking $\G=0, \F>0$, and the eigenvalues $\lambda$ are real.  For a constant electric field $\G=0, \F<0$ they are purely imaginary; this difference is the cause of the vacuum instability we will find for the constant electric field.

First, we consider a constant magnetic field, which we take to be along the 3 direction, $A = (0,0, -B x_1, 0)$, $B>0$, and we have $\G=0$, $\F=\frac{1}{2}B^2$, and

\begin{equation}
\label{eq:mageigen}
\frac{e}{2} \sigma_{\mu \nu} F^{\mu \nu}  \sim \diag(eB, eB, -eB, -eB)
\end{equation}
In this gauge the d'Alembertian $D^2$ becomes

\begin{eqnarray}
D^2 = p_0^2 - p_1^2 - (p_2 + e B x_1)^2 - p_3^2
\end{eqnarray}
and becomes after a unitary transformation
\begin{eqnarray}
D^2 = & e^{\imath p_1 p_2/e B} ( p_0^2 - p_1^2 - e^2 B^2  x_1^2 - p_3^2 ) e^{-\imath p_1 p_2/e B} 
\end{eqnarray}
where we have used the commutation relations $[ x_\mu, p_\nu ] = i g_{\mu \nu}$, and in particular \\
$[x_1, e^{a p_1}]~=~i a e^{a p_1}$. 

\begin{figure}[t]
\begin{center}
\epsfig{file=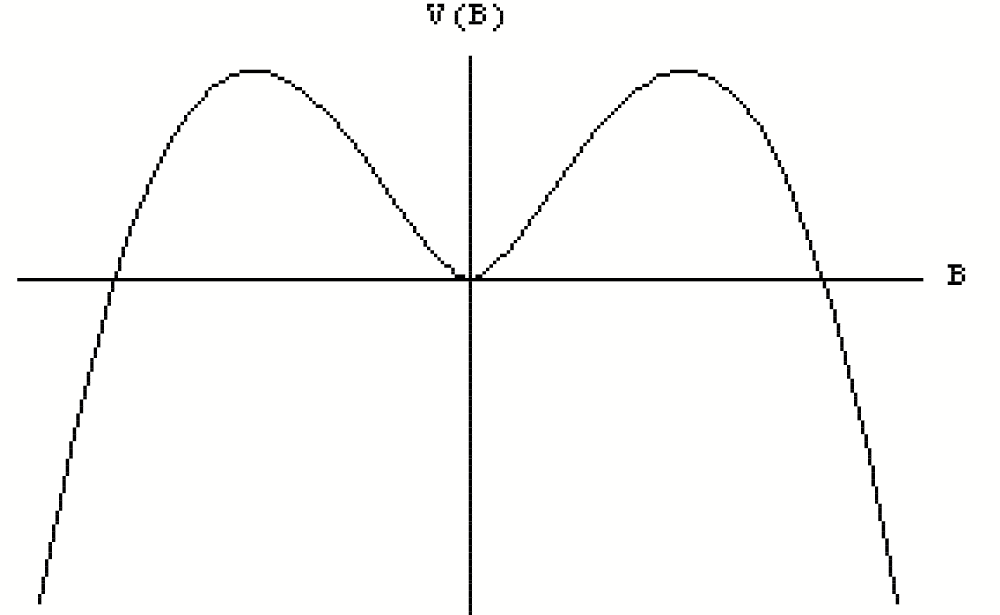}
\parbox{5.5in}{
\caption{1-loop effective potential for QED in a constant background magnetic field.  The apparent instability at large magnetic field strengths is an artifact of the 1-loop approximation.\label{fig:qed.pot}}}
\end{center}
\end{figure}

Therefore the 1-loop contribution to the effective Lagrangian is

\begin{equation}
\Lag^1 = -\frac{\imath}{2} \tr \log\left( e^{\imath p_2 p_1/e B} (p_0^2 - p_1^2 - e^2 B^2 x_1^2 - p_3^2) e^{-\imath p_2 p_1/eB} - \frac{e}{2} \sigma_{\mu \nu} F^{\mu \nu}\right)
\end{equation}
To evaluate this trace, we use the identity

\begin{equation}
\log(x) = \lim_{\epsilon \rightarrow 0} \frac{-\imath^\epsilon}{\Gamma(1+\epsilon)} \int_0^{\infty} dt\ t^{-1+\epsilon} e^{-\imath t x}
\end{equation}
This is related to the method used by Schwinger \cite{Schwinger:1951nm} (who introduced a lower cutoff into the integral instead of dimensionally continuing the argument), and amounts to rewriting the four-dimensional space-time loop momentum integral as the world-line integral of a particle moving in an external potential.  This is a close analogy of the world-sheet formalism of string theory; the world-line proper time parameter $t$ corresponds to a ``world-line modulus'' of the loop in the Feynman graph\footnote{The analogy between string theory and the Schwinger formulation of loop integrals was used in \cite{Dijkgraaf:2002xd} to calculate effective superpotentials in theories with $\N=1$ supersymmetry, by reducing a topological string theory calculation to a field theory calculation in Schwinger's formalism.  We will explain some key points of this work in section \ref{sec:susy.matter}.}.

\begin{eqnarray}
\Lag^1 &=& \frac{\imath^{1+\epsilon}}{2 \Gamma(1+\epsilon)}  \tr \int_0^{\infty} dt~t^{-1+\epsilon} e^{\imath p_2 p_1/eB} e^{-\imath t(p_0^2 - p1^2 - e^2 B^2 x_1^2 - p_3^2)} e^{-\imath p_2 p_1/eB} e^{\imath t \frac{e}{2} \sigma_{\mu \nu} F^{\mu \nu}} \nonumber \\
&=& \frac{\imath^{1+\epsilon}}{2 \Gamma(1+\epsilon)}  \int_0^{\infty} dt~t^{-1+\epsilon} \times \nonumber \\
&&\hspace{0.5in}2 \sum_{\lambda=\pm1} e^{\imath t e B \lambda} \langle x | e^{\imath p_2 p_1/eB} e^{-\imath t(p_0^2 - p_1^2 - e^2 B^2 x_1^2 - p_3^2)} e^{-\imath p_2 p_1/eB}  | x \rangle
\end{eqnarray}
For suitably large $\epsilon$ the integral converges, therefore this representation regulates the calculation.  In the second line we evaluated the trace over the anomalous magnetic moment operator using (\ref{eq:mageigen}), since the operator commutes with everything else in the trace.   The remaining trace may be evaluated as follows \cite{Itzykson:1980rh}

\begin{eqnarray}
&&\langle x | e^{\imath p_2 p_1/eB} e^{-\imath t(p_0^2 - p_1^2 - e^2 B^2 x_1^2 - p_3^2)} e^{-\imath p_2 p_1/eB}  | x \rangle \nonumber \\
&&\quad= \int d^4 p~ d^4 p'~\langle x |p \rangle \langle p |  e^{\imath p_2 p_1/eB} e^{-\imath t(p_0^2 - p_1^2 - e^2 B^2 x_1^2 - p_3^2)} e^{-\imath p_2 p_1/eB}  | p' \rangle \langle p' | x \rangle \nonumber \\
&&\quad= \int d^4 p~ d^4 p'~ \frac{e^{\imath (p-p').x}}{(2 \pi)^4}\langle p |  e^{\imath p_2 p_1/eB} e^{-\imath t(p_0^2 - p_1^2 - e^2 B^2 x_1^2 - p_3^2)} e^{-\imath p_2 p_1/eB}  | p' \rangle\nonumber\\
&&\quad= \int d^4 p~ d^4 p'~ \frac{e^{\imath (p-p').x}}{(2 \pi)} e^{\imath((p_2 p_1)/e B - (p_2' p_1')/eB)} e^{-\imath t (p_0^2 - p_3^2)} \nonumber \\
&&\hspace{2 in}\langle p_1 |  e^{-\imath -t( p_1^2 - e^2 B^2 x_1^2)}  | p_1' \rangle \delta^3((p-p')_{0,2,3}) \nonumber\\ 
&&\quad= \int d\omega~ d\omega'~ \frac{d^3p}{(2 \pi)^3}~ e^{- \imath t(p_0^2 - p_3^2)} e^{\imath(\omega - \omega') (x_1  + p_2/eB)} \langle \omega | e^{\imath t(p_1^2 + e^2 B^2 x_1^2)} | \omega' \rangle \nonumber \\
&&\quad= \frac{e B}{(2 \pi)^2 (\imath t)^{1/2}(-\imath t)^{1/2}} \int d\omega~ d\omega'~ \delta(\omega - \omega') \langle \omega | e^{\imath t(p_1^2 + e^2 B^2 x_1^2)} | \omega' \rangle \nonumber \\
&&\quad= \frac{e B}{(2 \pi)^2 t} \sum_{n=0}^{\infty} \exp(\imath t(n+\frac{1}{2}) 2 e B)
\end{eqnarray}
where we used the result for the energy levels of a harmonic oscillator

\begin{equation}
\tr \exp(\imath t(\frac{P^2}{2 m} + \frac{m \omega}{2} Q^2)) = \sum_{n=0}^\infty \exp(\imath t(n+\frac{1}{2}) \omega)
\end{equation}
Therefore the effective Lagrangian reduces to

\begin{eqnarray}
\Lag^1 &=& \frac{e B \imath^{1+\epsilon}}{8 \pi^2 \Gamma(1+\epsilon)} \int_0^\infty dt~t^{-2+\epsilon} \sum_{\lambda=\pm1} \exp(\imath e t B \lambda) \sum_{n=0}^\infty \exp(i e B t(2n+1)) \nonumber \\
&=& \frac{eB \imath^{1+\epsilon}}{8 \pi^{2}\Gamma(1+\epsilon)} (2 e B)^{1-\epsilon} \int_{0}^{\infty} dt~t^{-2+\epsilon} \frac{e^{-\imath t}+1}{1-e^{-\imath t}}
\end{eqnarray}
Rotating the integration contour $t \rightarrow \imath t$ we obtain
\begin{equation}
\Lag^1 = -\frac{e^{2}B^{2}}{4 \pi^{2}\Gamma(1+\epsilon)} (2 e B)^{-\epsilon} \int_{0}^{\infty} dt~t^{-2+\epsilon} \frac{e^{-t}+1}{1-e^{-t}}
\end{equation}
The integral may now be evaluated using the identity

\begin{equation}
\int_{0}^{\infty} dt\ t^{\sigma-1} \frac{e^{-\nu t} }{1-e^{-t}} = \Gamma(\sigma) \zeta(\sigma, \nu)
\end{equation}
where $\zeta(\sigma, \nu) = \sum_{n=0}^\infty (\nu + n)^{-\sigma}$ is the generalized Riemann zeta function.  Therefore

\begin{eqnarray}
\Lag^1 &=& -\frac{e^2 B^2}{4 \pi^2} \left(\frac{1}{2 e E}\right)^\epsilon \frac{-\Gamma(-1+\epsilon)}{\Gamma(1+\epsilon)} (\zeta(-1+ \epsilon, 0) + \zeta(-1+\epsilon, 1)) \nonumber \\
\end{eqnarray}
In taking the limit $\epsilon \rightarrow 0$, we renormalize the expression using a variant of the $\overline{MS}$ scheme \cite{Peskin:1995ev}\footnote{The difference is that we also subtract the $\log 2$ coming from the coefficient of $\Delta$}:

\begin{equation}
\frac{\Gamma(\epsilon)}{(4 \pi)^{2+\epsilon}} \left(\frac{1}{2 \Delta}\right)^{\epsilon} \rightarrow -\frac{1}{4 \pi^2} \log\left(\frac{\Delta}{\mu^2}\right)
\end{equation}
and use the property of the $\zeta$-function
\begin{equation}
\zeta(-m, \nu) = - \frac{B_{m+1}(\nu)}{m+1}
\end{equation}
where $m=0, 1, \ldots$, and $B_{m+1}(\nu)$ are the Bernoulli polynomials, in particular \\
$B_2(x) = x^2 -x + 1/6$.  Putting this all together, we find for the effective potential

\begin{eqnarray}
\label{eq:qed.veff}
\Veff =  \frac{B^2}{2} - \frac{e^2 B^2}{24 \pi^2} \log(e B/\mu^2)
&=& \frac{B^2}{2} - \frac{b_0 B^2}{2 e} \log(e B/\mu^2)
\end{eqnarray}
where we recognize the 1-loop QED $\beta$-function coefficient $b_0 = \frac{e^3}{12 \pi^2}$.  This potential is plotted in figure \ref{fig:qed.pot}.

For small external fields $e B < \mu^2$ the second term is positive, and the effective potential has a local minimum at $B=0$.  At larger field strengths there appears to be a local maximum and the potential eventually becomes arbitrarily negative.  However, in precisely this limit the 1-loop approximation breaks down, because the quantum correction term dominates and is no longer small compared to the classical term.  Therefore, for large enough magnetic fields one needs to also consider the higher loop corrections.

We turn now to the electric case.  Using the form of $\F = \frac{1}{2}(B^2 -E^2)$, we may obtain the effective potential for a constant background electric field $E \neq 0, B = 0$ by formally continuing $B \rightarrow \imath B \equiv E$ in (\ref{eq:qed.veff}). This introduces a factor of $\imath$ into the argument of the logarithm, and therefore the effective Lagrangian in a background electric field is complex.

Since the amplitude for a vacuum in the far past to remain in the far future is given by

\begin{equation}
\langle 0_{+} | 0_{-} \rangle = e^{\imath \Gamma}
\end{equation}
the probability of vacuum decay, per unit time and volume, is given by 
\begin{equation}
2 \im \Lag = -\im\frac{e^{2}E^{2}}{12 \pi^{2}} \log(\imath)= \frac{e^{2}E^{2}}{24 \pi}
\end{equation}
and the constant electric field background is unstable against pair production of positron-electron pairs.

The result for a non-zero electron mass can also be computed following the above steps, and one finds

\begin{equation}
2 \im \Lag = \frac{e^{2}E^{2}}{4 \pi^{3}} \sum_{N=1}^{\infty} \frac{1}{N^{2}} \exp(\frac{-N \pi m^{2}}{e E})
\end{equation}
which is non-perturbative in the RG-invariant field combination $eE$.  Again we see that the background field method produces non-perturbative information from a perturbative calculation.

In a general constant background with $\F \neq 0, \G \neq 0$, the effective Lagrangian is that of Euler and Heisenberg \cite{Heisenberg:1936qt}, which takes the form (before regularization and renormalization)

\begin{equation}
\label{eq:eh}
\Lag^{1} = \frac{1}{8 \pi^{2}} \int_{0}^{\infty} dt~t^{-1} e^{\imath s m^{2}}\left( e^{2} a b \frac{\cosh(e a t) \cos(e b t)}{\sinh(e a t) \sin(e b t)} \right)
\end{equation}
where $a^{2} - b^{2} = E^{2}-B^{2}$, $a b = E \cdot B$.  A list of references to recent work on this Lagrangian and related matters may be found in \cite{Dunne:2003tr}.

\subsection{Yang-Mills theory}
\label{sec:ym}

To calculate the 1-loop effective action for four-dimensional Yang-Mills theory we again use the background field method.  This calculation and related results were developed by a number of authors, including \cite{Duff:1975ue,Batalin:1977uv,Matinyan:1978mp,Nielsen:1978rm,Pagels:1978dd,Brown:1979bv,Yildiz:1980vv,Flyvbjerg:1980qv,Flyvbjerg:1981rs,Jezabek:1981bs}.

The Yang-Mills Lagrangian is

\begin{equation}
{\cal L} =  -\frac{1}{4} F_{\mu \nu}^a F^{\mu \nu a}
\end{equation}
We split the gauge field into a classical background field $A$ and a fluctuating quantum field $a$:

\begin{equation}
A_\mu^a(x) \rightarrow A_\mu^a(x) + a_\mu^a(x)
\end{equation}
The covariant derivative $(D_\mu)^{ac}=\partial_{\mu} \delta^{ac} + \imath g f^{a b c}A_{\mu}^{b}$ is defined with respect to the background gauge field, and we will integrate over the quantum field $a$ in the path integral.  Then the field strength becomes

\begin{equation}
F_{\mu \nu}^a \rightarrow F_{\mu \nu}^a + D_\mu a_\nu^a - D_\nu a_\mu^a + \imath g f^{a b c} a_\mu^b a_\nu^c
\end{equation}
In background gauge $D_{\mu} A^{\mu a} = 0$, the gauge-fixed Lagrangian is
 
\begin{eqnarray}
\label{eq:lag.ym}
{\cal L} &=&  -\frac{1}{4} (F_{\mu \nu}^a + D_{\mu} a_{\nu}^{a} - D_{\nu} a_{\mu}^{a} +\imath g f^{a b c} a_{\mu}^{b} a_{\nu}^{c})^{2} \nonumber \\
&&\hspace{0.5in}-\frac{1}{2} (D^\mu a^{\mu a})^2 + \overline c^a(-(D^2)^{a c}  -\imath g D^{\mu a d} f^{d b c} a_\mu^b) c^c
\end{eqnarray}
where $c, \overline c$ are the Faddeev-Popov ghosts corresponding to the gauge fixing.

As before, the effective action to 1-loop order is given by evaluating the path integral 
\begin{equation}
e^{\imath\Gamma[A]} = \int {\cal D} a {\cal D} c {\cal D} \overline c e^{\imath \int d^4x {\cal L}}
\end{equation}
to quadratic order in the fluctuations.  Expanding (\ref{eq:lag.ym}) to quadratic order, we find

\begin{equation}
{\cal L}_{\hbox{\it quad}} = -\frac{1}{2} a_\mu^a \left[ (-D^2)^{a b} g^{\mu \nu} - 2 \imath g F^{\mu \nu c} f^{c a b} \right] a_{\mu}^b + \overline c^{a} \left[ -(D^2)^{ab} \right] c^b
\end{equation}
As in QED, the new interaction term $- 2 \imath g F^{\mu \nu c} f^{c a b}a^{\mu a} a^{\nu b}$ is an anomalous magnetic moment interaction of two spin-1 gluons with the background field $F^{\mu \nu c}$.  Introducing the generator of spin-1 Lorentz transformations
\begin{equation}
(J^{\rho \sigma})_{\alpha \beta} = \imath(\delta^{\rho}_{\alpha} \delta^{\sigma}_{\beta} - \delta^{\sigma}_{\alpha} \delta^{\rho}_{\beta})
\end{equation}
the operator $- 2 \imath g F^{\mu \nu c} f^{c a b}$ can be rewritten as $-2 \imath (\frac{1}{2} F^{c}_{\rho \sigma} J^{\rho \sigma})^{\mu \nu} f^{c a b}$, emphasizing the similarity to the operator (\ref{eq:qed.det}) for spin-$\frac{1}{2}$ electrons in QED.  The spin interaction for the ghost fields vanishes since they have spin 0.

Therefore the path integral to 1-loop order is Gaussian and can be evaluated, giving the 1-loop effective Lagrangian

\begin{equation}
{\cal L}_{\mbox{\it \!eff}} = -\frac{1}{4} F_{\mu \nu}^a F^{\mu \nu a} + \frac{\imath}{2} \log \det((-D^2)^{a b} g^{mu \nu} - 2 \imath g F^{\mu \nu c} f^{c a b}) - \imath \log \det((-D^2)^{a b}))
\end{equation}
We can evaluate these determinants by restricting to covariantly constant fluctuations of the gauge fields:

\begin{eqnarray}
D_\rho F^{\mu \nu} = 0
\Leftrightarrow [ D_\rho, F^{\mu \nu} ] = 0 
\end{eqnarray}
where we write the field strength as a matrix in colour space $(F_{\mu \nu})^{a b} = f^{a b c} F_{\mu \nu}^c$, and the second form follows because the covariant derivative in the adjoint representation acts by matrix commutation.  Using the Jacobi identity $[D_\sigma, [D_\rho, F_{\mu \nu}]] + \hbox{\it perm.} = 0$ it follows that

\begin{equation}
[F_{\mu \nu},F_{\rho \sigma}] = 0
\end{equation}
i.e.~the colour matrices $F_{\mu \nu}$ form a commuting set and may be simultaneously diagonalized.  In other words, by a gauge transformation we may rotate a given gauge field configuration into the Cartan subalgebra.  Then

\begin{eqnarray}
\label{eq:ym.color}
{\cal L}^{1} &=& \frac{\imath}{2}\tr \sum_{\alpha} \log(-D^{(\alpha)2} g^{\mu \nu} - 2 \imath g F^{\mu \nu {(\alpha)}}) - \imath \sum_{\alpha} \tr \log((-D^{(\alpha)})^{2}) \nonumber \\
&=& \imath\tr \sum_{\alpha>0} \log(-D^{(\alpha)2} g^{\mu \nu} - 2 \imath g F^{\mu \nu (\alpha)}) - 2 \imath \sum_{\alpha>0} \tr \log((-D^{(\alpha)})^{2})
\end{eqnarray}
where the sum is over the positive roots $\alpha$ of ${\cal G}$.  In the second line we used that each root $\alpha$ is paired with a negative root $-\alpha$, and the zero roots do not contribute.  We also defined effective quantities

\begin{eqnarray}
D_{\mu}^{(\alpha)} &=& \partial_\mu + i g {\alpha}_j A_{\mu}^j \nonumber \\
F_{\mu \nu}^{(\alpha)}&=& \alpha_{j} F_{\mu \nu}^{j}  \nonumber \\
A_{\mu}^{(\alpha)} &=& \alpha_j A_{\mu}^j
\end{eqnarray}
in terms of the simple roots $(\alpha_1, \ldots, \alpha_r)$, $r=\hbox{rank}({\cal G})$, which span root space.

In other words, we have reduced the computation of the 1-loop effective action for a non-abelian gauge group ${\cal G}$ to that of an Abelian $\U(1)^r$ gauge theory, where the $j$'th ``photon'' carries charges $g \alpha_{j}$ with respect to the different $\U(1)$ gauge factors.  The situation is therefore quite similar to that of QED, which we studied in the previous section, except there is more than one type of ``electromagnetic field'', and the charged particles are spin-1 photons, not spin-$\frac{1}{2}$ electrons.

At this point we need to choose the orientation for the effective $\U(1)$ gauge fields in four-dimensional space; when the rank of the gauge group is larger than 1, the ``electromagnetic fields'' may point in different spatial directions.  Most of the early work on this problem either considered $\SU(2)$ \cite{Batalin:1977uv,Matinyan:1978mp}, or chose to align all the effective $\U(1)$ gauge fields parallel \cite{Ambjorn:1979ff}. However, it was shown in subsequent work that for $2 < N \le 4$ the lowest-energy configuration is to choose the fields to be mutually orthogonal \cite{Flyvbjerg:1980qv}.  For $N \ge 4$, i.e.~rank higher than 3, it is no longer possible to choose all vectors to be orthogonal in three-dimensional space, and for $N\rightarrow \infty$ the minimum energy configuration corresponds to an isotropic distribution in space \cite{Flyvbjerg:1981rs,Jezabek:1981bs}.

For simplicity, we will henceforth restrict to the $\SU(2)$ case.  The essential features are seen in this case; in particular we will see that {\it any} choice of covariantly constant field strength gives rise to a vacuum instability, and therefore the 1-loop result is at best only an approximation to the true vacuum.  This instability persists for the non-parallel gauge field orientations mentioned above.

We can now proceed as in section \ref{sec:qed}.  Again taking a constant magnetic field, the eigenvalues of $-2 \imath g F_{\mu \nu}$ are $(\pm 2 g B, 0, 0)$.  The two zero eigenvalues cancel with the contribution from the ghost determinant in (\ref{eq:ym.color}), giving

\begin{equation}
\Lag^{1} = \sum_{\lambda = \pm 1} \tr \log(-D^{2} - 2 \lambda B)
\end{equation}

After manipulations similar to QED, we find

\begin{eqnarray}
\Lag^{1} &=& - \frac{g B }{8 \pi^{2}} (g B)^{-\epsilon}\frac{\imath^{1+\epsilon}}{\Gamma(1+\epsilon)} \int_{0}^{\infty}dt~t^{-2+\epsilon} \times \sum_{\lambda=\pm 1}\exp(-\imath t(1-2 \lambda)) \nonumber \\
&&\hspace{0.5in} \times \sum_{N=0}^{\infty} \exp(-2 \imath t N)
\end{eqnarray}
Note that we can no longer unconditionally rotate the contour by taking $t \rightarrow - \imath t$, because the mode with $(\lambda, N) = (1,0)$ would diverge like $e^{t}$.  This is the {\it unstable mode} found by Nielsen and Olesen \cite{Nielsen:1978rm}, which will give rise to an imaginary part for the 1-loop effective Lagrangian even in the magnetic case.  To proceed, we subtract and add the $(\lambda, N) = (1,0)$ term:

\begin{eqnarray}
\Lag^{1} &=& - \frac{g B }{8 \pi^{2}}(g B)^{-\epsilon} \frac{1}{\Gamma(1+\epsilon)} \left\{ \imath^{1+\epsilon}\int_{0}^{\infty}dt~t^{-2+\epsilon} \frac{e^{\imath t} + e^{-3 \imath t}}{1-e^{-\imath t}} - e^{\imath t} \nonumber \right.\\
&&\hspace{0.5in}\left.+ \imath^{1+\epsilon}\int_{0}^{\infty}dt~t^{-2+\epsilon} e^{\imath t}\right\} \nonumber \\
&=& - \frac{g B}{8 \pi^{2}}(g B)^{-\epsilon} \frac{1}{\Gamma(1+\epsilon)} \left\{ \imath^{1+\epsilon}\int_{0}^{\infty}dt~t^{-2+\epsilon} \frac{e^{-\imath t} + e^{-3 \imath t}}{1-e^{-\imath t}}\right. \nonumber \\
&&\hspace{0.5in}\left.+ \imath^{1+\epsilon}\int_{0}^{\infty}dt~t^{-2+\epsilon} e^{\imath t}\right\} \nonumber \\
&=& - \frac{g B}{8 \pi^{2}}(g B)^{-\epsilon} \frac{1}{\Gamma(1+\epsilon)} \left\{ \imath^{1+\epsilon} \left(\frac{-\imath}{2}\right)^{-1+\epsilon} \int_{0}^{\infty}dt~t^{-2+\epsilon} \frac{e^{-t} + e^{-3 t}}{1-e^{-t}} \right.\nonumber \\
&&\hspace{0.5in}\left.+ \imath^{1+\epsilon} \imath^{-1+\epsilon} \int_{0}^{\infty}dt~t^{-2+\epsilon} e^{- t}\right\} \nonumber \\
&=& - \frac{g B}{8 \pi^{2}}(g B)^{-\epsilon} \frac{1}{\Gamma(1+\epsilon)} \left\{ -2\left(\frac{1}{2}\right)^{\epsilon} \int_{0}^{\infty}dt~t^{-2+\epsilon} \frac{e^{-t} + e^{-3 t}}{1-e^{-t}} \right.\nonumber \\
&&\hspace{0.5in}\left.+ (-1)^{\epsilon} \int_{0}^{\infty}dt~t^{-2+\epsilon} e^{- t}\right\}
\end{eqnarray}
where we rotated the two integration contours by $t \rightarrow -\imath t$, $t \rightarrow \imath t$ respectively.  The integrals may now be evaluated in terms of zeta functions, giving

\begin{eqnarray}
\Lag^{1} &=& -\frac{(g B)^{2}}{8 \pi^{2}} \frac{\Gamma(\epsilon)}{(-1+\epsilon) \Gamma(1+\epsilon)} \left[ \left(\frac{1}{2 g B}\right)^{\epsilon}(-2) (\zeta(-1+\epsilon, \frac{1}{2}) \right.\nonumber \\
&&\hspace{0.5in}\left.+ \zeta(-1+\epsilon, \frac{3}{2})) + \left(\frac{-1}{g B}\right)^{\epsilon}\right] \nonumber \\
&=& -\frac{(g B)^{2}}{8 \pi^{2}} \log(g B/\mu^{2}) \left[ -2(\zeta(-1,\frac{1}{2}) + \zeta(-1,\frac{3}{2})) + 1\right] \nonumber \\
&&\hspace{0.5in}- \frac{(g B)^{2}}{8 \pi^{2}} \log(-1) \nonumber \\
&=& - \frac{11}{6}\frac{(g B)^{2}}{8 \pi^{2}} \log(g B/\mu^{2}) + \imath \frac{(g B)^{2}}{8 \pi} \nonumber \\
&=& + \frac{\beta(g)}{2 g} B^{2} \log(g B/\mu^{2}) + \imath \frac{(g B)^{2}}{8 \pi}
\end{eqnarray}
where in the last line we recognized the 1-loop $\beta$-function coefficient.  As before the pure electric field result may be obtained by analytic continuation. If we consider a background with $\G \neq 0$, then the effective Lagrangian will be a generalization of the Euler-Heisenberg Lagrangian (\ref{eq:eh}) \cite{Batalin:1977uv,Matinyan:1978mp}.  In all cases the background is unstable, in contrast to QED, for which only the electric background is unstable.

Note that because of asymptotic freedom the sign of the 1-loop term is opposite to that of  QED (\ref{eq:qed.veff}); therefore the effective potential has a similar form to figure \ref{fig:gn.pot}.  The lesson we can draw from this analysis is that the ``perturbative vacuum'', where we consider excitations around the zero-field background, is an unstable field configuration.  The Yang-Mills vacuum lowers its energy by spontaneously generating a non-zero background field.  This can be seen as a vacuum anti-screening effect by the gluons, which are charged under the gauge group and can act as sources for other gluons.  Turning on a covariantly constant background field indeed lowers the vacuum energy, but this field configuration is itself unstable (not to mention violating Lorentz invariance), so the ``true'' vacuum is some other field configuration.  An ansatz for the vacuum (the ``Copenhagen vacuum'') was proposed in \cite{Nielsen:1979tr}, based on exciting the unstable mode of the constant-field vacuum.

The background field method is non-perturbative in the background field (since it is not used as an expansion parameter), which allowed us to make some progress, but excitations around this field still must be calculated perturbatively.  This means that we can only trust our 1-loop calculation when the effective coupling constant is small, however this is counteracted by the negative sign of the 1-loop $\beta$-function, which tells us that $g$ will grow towards the IR.  

Explicitly, to 1-loop order the running of the Yang-Mills coupling constant is given by

\begin{equation}
g_{\hbox{\it eff\,}}^{2}(q) = \frac{g^{2}}{1+\frac{11 g^{2}}{96 \pi^{2} N} \log(q/\mu)}
\end{equation}
which diverges at the finite energy scale

\begin{equation}
q = \mu \exp(-\frac{96 \pi^{2} N}{11 g^{2}}) \equiv \Lambda_{\hbox{\small YM}}
\end{equation}
Therefore, we can not trust our 1-loop effective potential at energies comparable to or lower than $\Lambda_{\mbox{\small YM}}$.  Nevertheless, it is expected (based on lattice simulations and other theoretical work) that the qualitative picture remains true, and the vacuum of Yang-Mills theory is associated to non-trivial gauge field backgrounds, which give rise to confinement, generation of a mass gap (the appearance of massive glueballs in the spectrum replacing the massless gluons), and other poorly-understood low-energy physics.

\subsection{Constraints on the effective potential from the trace anomaly}
\label{sec:ym.anom}

We have seen that the effective potential of quantum field theories must be consistent with the trace anomaly, in particular it satisfies the Callan-Symanzik equation.  Once we have calculated the quantities $\beta$ and $\gamma$ for a particular theory, we can use the Callan-Symanzik equation to constrain the possible form of corrections to the classical potential in an $\it arbitrary$ field background.

For $\SU(2)$ Yang-Mills theory we found that the effective potential in a covariantly constant field background with $\F = \frac{1}{4} F_{\mu \nu}^{a} F^{\mu \nu a} \neq 0$, $\G = \frac{1}{4} F_{\mu \nu}^{a} \tilde{F}^{\mu \nu a} = 0$ is (suppressing the trace over the colour indices):

\begin{equation}
\label{eq:ym.veff}
\Veff = \frac{1}{4} F^{2} + \frac{11 g^{2}}{16 \times 48 \pi^{2} N} F^{2} \log(g^{2} F^{2}/\mu^{4})
\end{equation}
Applying the Callan-Symanzik equation

\begin{equation}
\label{eq:cs2}
\left[ \mu \frac{\partial}{\partial \mu} + \beta(g) \frac{\partial}{\partial g} - \gamma F \frac{\partial}{\partial F} \right] \Veff = 0
\end{equation}
we find that $\beta = \gamma g = -\frac{11g^{3}}{3 (4 \pi)^{2} N}$.  These are properties of the Lagrangian, and do not depend on the particular background we evaluate it in; moreover to 1-loop order they are independent of the renormalization scheme.  We now look for more general functions $V$ that solve (\ref{eq:cs2}), to see what possible corrections may appear in other field backgrounds.  The equation (\ref{eq:cs2}) can be solved by a series of the form

\begin{equation}
V = \sum_{i=0}^{\infty}  a_{i}(g) F^{2} \log(g F/\mu^{2})^{i}
\end{equation}
where the $a_{i}(g)$ satisfy a set of coupled differential relations of the form

\begin{equation}
\label{eq:ym.diffrel}
\gamma g \frac{d a_{i}}{d g} - 2 \gamma a_{i} + (i+1) a_{i+1} = 0
\end{equation}
where we have used the relation $\beta = \gamma g$ that we found above.

If we assume that to 1-loop order, the correction series in a particular background terminates at some order $k$, then we can integrate the relations 
 (\ref{eq:ym.diffrel}) and impose that the function $V$ reduces to the classical potential $V = \frac{1}{4}F^{2}$ plus corrections that are higher powers of $g$.  We find
 
\begin{eqnarray} 
a_{k} &=& 0 \nonumber \\
a_{k-1} &=& C_1 g^{2}\nonumber \\ 
a_{k-2} &=& C_2 g^{2} - \frac{(k-1) C_1}{\alpha} \nonumber \\
a_{k-3} &=& \ldots
\end{eqnarray}
where we define the 1-loop $\beta$ function $\beta(g) = \alpha g^{3}$, $\alpha = -\frac{11}{3 (4 \pi)^{2} N}$.  Thus, consistency with tree level fixes $k=2$ and the value of $C_1$, and subject to the assumptions above, the general effective potential for a (not necessarily constant) background with $\F \neq 0, \G = 0$ is

\begin{eqnarray}
\label{eq:ym.univ}
V &=& \frac{1}{4} F^{2} + C_2 g^{2} F^{2} - \frac{\alpha g^{2}}{8} F^{2} \log(g^{2} F^{2}/\mu^{4}) \nonumber \\
&=& \frac{1}{4} F^{2} + C_2 g^{2} F^{2} + \frac{11 g^{2}}{8 \times3 (4 \pi)^{2} N } F^{2} \log(g^{2} F^{2}/\mu^{4})
\end{eqnarray}
The unfixed constant $C_2$ reflects the ability to shift the arbitrary renormalization scale $\mu$, as well as the possible instability of the field background if $C_2$ is complex.  Similar arguments constrain the form of $V$ in an arbitrary background with $\G \ne 0$, which gives a generalization of the Euler-Heisenberg Lagrangian \cite{Batalin:1977uv}. Note in particular that the sign of the 1-loop contribution -- and therefore the existence of the unstable perturbative vacuum -- depends on the negative sign of $\beta(g)$.

Note that this method does not rely on knowledge of the precise the form of $F_{\mu \nu}^{a}$ in 4-dimensional space-time, or in the internal (colour) space.  Non-constant field configurations may have complicated derivative terms in their effective Lagrangian, but for configurations that satisfy our assumptions, the trace anomaly constrains the non-derivative terms to reduce essentially to the form of the constant field result obtained above. 
However, as noted above this does not allow us to reliably estimate the vacuum expectation value $\langle F^2 \rangle$, because the 1-loop approximation still breaks down before we reach the dynamical scale $\Lambda$ characteristic of confinement\footnote{A more reliable estimate of $\langle F^2 \rangle$ for QCD was made by Shifman et.~al.~\cite{Shifman:1979bx} using charmonium sum rules.}.

In section \ref{sec:susy} we will turn this argument around, and use 1-loop anomalies to compute the effective superpotential of $\N=1$ supersymmetric Yang-Mills theory directly.  The 1-loop anomaly calculation is exact in supersymmetric theories, which allows us to find the exact effective superpotential without needing to perform an explicit path integral calculation around the vacuum field configuration.  Indeed, the precise nature of the $\N=1$ vacuum is unknown, although we can compute some of its properties exactly.

\section{$\N=1$ supersymmetric gauge theories}
\label{sec:susy}

In a supersymmetric theory, the Lagrangian may contain terms of the form

\begin{equation}
\int d^{2} \theta~W(\Phi_{i}) + h.c.
\end{equation}
where the integral is over half of superspace, and $W$ is the {\it superpotential} of the theory.  It has dimension 3 and is a function of the chiral superfields $\Phi_{i}$ and not of their antichiral hermitian conjugates $\overline \Phi_{i}$.  The supersymmetric vacua of the theory are determined by the ``F-term'' constraints

\begin{equation}
\frac{\partial W}{\partial \Phi_i} = 0
\end{equation}
modulo complexified gauge transformations.  In terms of the superpotential, the ordinary bosonic potential of the theory is given by

\begin{equation}
V(\phi_i) = \sum_i | \frac{\partial W}{\partial \phi_i}|^2 + \frac{g^2}{2} (D^a)^2
\end{equation}
where $\phi_i$ are the lowest components of the chiral superfields $\Phi_i$ and $D^a = \sum_i | \phi_i | ^2 t^a$, where $t^a$ are the generators of the gauge group.

There are two key results that allow us to compute the effective superpotential exactly in many supersymmetric theories: in a {\it Wilsonian} approach where we integrate over loop momenta down to a momentum cutoff, the superpotential only receives one-loop and non-perturbative corrections; and it is a holomorphic function of the chiral superfields and coupling constants.  The meaning of these statements is somewhat subtle, and bears further explaining.

Until now, we have considered the effective potential defined by the non-derivative terms in the generating functional of 1-particle irreducible (1PI) diagrams of the theory that is obtained by integrating over the fluctuating fields.  We found that in four-dimensional gauge theories this object receives contributions to all loop orders in perturbation theory, corresponding to Feynman diagrams in the background field with arbitrarily many internal loops.  This remains true in a supersymmetric theory.  Moreover, higher loop corrections will generically not be holomorphic.

The Wilsonian approach to the effective action is to integrate over all loop momenta down to some cutoff scale; the resulting functional depends on the lower-momentum modes but has no dependence on momenta higher than the cutoff.  If we integrate all the way to zero momentum we would recover the 1PI generating functional.  In supersymmetric gauge theories, Shifman and Vainshtein \cite{Shifman:1986zi} showed that the 2-loop and higher contributions are infrared effects; they only enter the Wilsonian effective action as the cutoff is taken to zero, and in computing matrix elements of Wilsonian quantities (averaging them over the external fields).  For finite cutoff, the terms appearing in the Wilsonian effective action arise only from tree-level and 1-loop contributions.

It is important to note that the parameters (fields, coupling constants) that appear in the Wilsonian effective action are not the physical quantities that would be measured in an experiment; indeed, the latter receive corrections to all orders.  It would appear that the Wilsonian approach is missing the effects of the higher-loop contributions; as we saw in non-supersymmetric Yang-Mills theory the higher loop corrections are vital for understanding the vacuum structure, because they dominate at low energies.

The resolution, emphasized by \cite{Shifman:1991dz,Dine:1994su}, is that the all-loop, non-holomorphic 1PI effective superpotential may be brought into the 1-loop, holomorphic Wilsonian form by a suitable (non-holomorphic, field- and coupling- dependent) change of variable.  In other words, the 1PI effective superpotential is {\it resummed} into the Wilsonian form by this change of variable.  This means that in supersymmetric theories the higher order corrections to the effective superpotential arising from the trace anomaly must all be related to the form of the 1-loop term, written in different variables.  For example, in $\N=1$ supersymmetric Yang-Mills theory this is intimately related to the existence of the exact NSVZ $\beta$-function \cite{Novikov:1983uc}, which has the form of a geometric series.

Therefore, for supersymmetric theories we can confidently use the 1-loop Wilsonian effective potential to study the theory beyond the range where 1-loop perturbation theory naively breaks down, because we know that written in terms of physical quantities the 1-loop calculation sums the contributions to all loop orders.  If in addition the non-perturbative corrections to the effective superpotential are calculable (by holomorphy and symmetry constraints, this is often the case), then we can obtain the exact effective superpotential, and by extension, exact results about the vacuum of the theory.  The price is that to rewrite these exact Wilsonian results in terms of physical quantities one must undo the complicated change of variables.

\subsection{$\N=1$ Yang-Mills}
\label{sec:susy.ym}

The effective superpotential for $\N=1$ Yang-Mills was constructed in \cite{Veneziano:1982ah}, by writing an effective Lagrangian whose symmetry transformations reproduced the correct 1-loop anomalies.  This is essentially the approach we used in earlier sections.

The Lagrangian for $\N=1$ Yang-Mills theory is:

\begin{eqnarray}
{\cal L} =-\frac{1}{4g^{2}} F_{\mu \nu}^{a} F^{\mu \nu a} + \theta F_{\mu \nu}^{a} \tilde{F}^{\mu \nu a}+  \frac{\imath}{2} \overline{\lambda}^{a} \dslash D_{a b} \lambda^{b} + \ldots
\label{sym}
\end{eqnarray}
where we have suppressed the gauge-fixing, ghost and auxilliary terms.  In superfield notation this can be written as

\begin{equation}
\Lag = -\int d^2 \theta~\frac{1}{4 g^{2}} \tr W_{\alpha} W^{\alpha}  + h.c. = \int d^2 \theta~\tau  S + h.c.
\end{equation}
where we define

\begin{eqnarray}
S&=& -\frac{1}{32 \pi^{2}} \tr W_{\alpha}^{2} \nonumber \\
\tau &=& \frac{8 \pi^{2}}{g^{2}} + \imath \theta
\end{eqnarray}
$S$ is the ``gaugino bilinear superfield'', whose lowest component is $\tr \lambda_{\alpha}^{2}$.  In particular, $S$ and $\tau$ are both complex.

The expansion of the composite superfield $S$ in terms of component fields includes a term $\tr (F_{\mu \nu}^{a})^2$ quadratic in the Yang-Mills field-strength tensors, which one might be tempted to identify with a scalar ``glueball'' operator of the Yang-Mills theory.  However, $S$ cannot be interpreted as a dynamical glueball superfield, because the Yang-Mills field-strengths appear as auxilliary fields in $S$ and are therefore non-dynamical \cite{Sannino:2003xe}.  The approach of studying the vacuum of $\N=1$ Yang-Mills theory by introducing a non-dynamical composite field is essentially the same approach we took in probing the Gross-Neveu model for the existence of a symmetry-breaking fermion condensate; here we are probing for a gaugino condensate, to which we associate the composite field $S$ that includes the gaugino bilinear.
In this sense, the effective superpotential $W(S)$ we will obtain is part of a ``minimal Lagrangian'' that describes the symmetries and anomalies of the theory, but is not an effective Lagrangian for physical degrees of freedom.  In particular, upon extremizing the effective superpotential $W(S)$ we will obtain the value of the gaugino condensate in the vacua of $\N=1$ Yang-Mills.

As before, the Callan-Symanzik equation constrains the form of corrections arising from the anomalous breaking of scale-invariance\footnote{In $\N=1$ Yang-Mills theory the trace anomaly is part of an anomaly multiplet that also includes the axial anomaly, and a superconformal anomaly.  By supersymmetry, the constraints from the other anomalies are equivalent to that of the trace anomaly.}:

\begin{equation}
\left[\gamma S \frac{\partial}{\partial S}- \beta(g) \frac{\partial}{\partial g} - \mu \frac{\partial}{\partial \mu} \right]  \Weff(S) = 0
\end{equation}
As we have seen in previous examples, it can be solved by a function of the form

\begin{equation}
\Weff(S) = \frac{C_1}{g^{2}} S + C_2 S + C_3 S \log(S/\mu^{3})
\end{equation}
and we find $\gamma = 0$, $C_1=8 \pi^{2}$, $C_3 = \frac{16 \pi^{2} \beta(g)}{3 g^{3}} = N$, where $\beta(g) = -\frac{3N g^{3}}{(4 \pi)^{2}}$ to 1 loop.  Therefore

\begin{eqnarray}
\Weff(S)&=& \tau S + C_2 S + N S \log(\frac{S}{\mu^{3}}) \nonumber \\
&=& C_2 S + N S\log(S/\Lambda^{3})
\end{eqnarray}
where we introduced the dynamical scale $\Lambda$ via the running coupling relation

\begin{equation}
\label{eq:runningcoupling}
\tau(\mu) - 3 N \log \mu = 3 N \log \Lambda 
\end{equation}

As in other examples, the constant $C_2$ is not fixed by symmetries and may depend on the renormalization scheme.  A value can be fixed following the approach of \cite{Cachazo:2002ry}.  Using an instanton calculation \cite{Novikov:1983uc}, the value of the gaugino condensate can be obtained directly, giving rise to the value of the superpotential in the vacuum:

\begin{equation}
\label{eq:sym.vac}
\Weff(\Lambda)= N (\Lambda^{3 N})^{1/N}
\end{equation}
The field $S$ can be introduced by performing a Legendre transformation

\begin{equation}
\Weff(\Lambda, C, S) = N C^{3} + S \log(\frac{\Lambda^{3 N}}{C^{3 N}})
\end{equation}
Integrating out $S$ recovers the previous expression (\ref{eq:sym.vac}).  If instead we integrate out $C$, then we recover the Veneziano-Yankielowicz superpotential

\begin{equation}
\Weff(S, \Lambda) = N S( \log(\frac{S}{\Lambda^{3}}) - 1)
\end{equation}
which fixes the constant $C_2 = -N$.

Since the field $S$ here is complex, the F-term constraint $\frac{\partial W}{\partial S}=0$ gives $N$ distinct vacua (related by a phase, i.e.~vacuum angle $\theta$)

\begin{equation}
\label{eq:gauginocond}
\langle S \rangle = e^{2 \pi \imath k/N_c} \Lambda^3\quad\quad k=0, \ldots, N_c-1
\end{equation}

Furthermore, as noted in the previous section, this Wilsonian effective superpotential does not receive corrections beyond one loop.  Therefore the vacuum expectation value $\langle S \rangle \propto \langle \tr W_\alpha W^\alpha \rangle$ is exact, and the $N$ vacua of $\N=1$ supersymmetric $\SU(N)$ Yang-Mills theory have a non-vanishing gaugino condensate.

Note that the Callan-Symanzik anomaly calculation does not assume a particular form of the background gauge field configuration.  A covariantly constant background field strength was considered in \cite{Kay:1983mh}, generalizing the Yang-Mills calculations reviewed in section \ref{sec:ym}.  As in the non-supersymmetric case, a constant background field strength causes the vacuum energy to decrease, but there is still an instability at the 1-loop level\footnote{This is not surprising since this field configuration is not supersymmetric.}.  A field theoretical derivation of the Veneziano-Yankielowicz superpotential is not known - this would amount to knowing the field configuration in the $\N=1$ Yang-Mills vacuum and integrating over the fluctuations around this background.

\section{$\N=1$ theories with matter}
\label{sec:susy.matter}

One of the starting-points for the recent work on $\N=1$ gauge theories with adjoint matter was the conjecture \cite{Dijkgraaf:2002fc,Dijkgraaf:2002vw,Dijkgraaf:2002dh} that the effective superpotential is computed by an associated bosonic large-$N$ matrix integral, which may be evaluated by counting planar diagrams.  This conjecture comes from string theory, and follows a chain of reasoning that is the culmination of extensive research on the relationship between string theory and gauge theories.

The steps in the conjecture can be summarized as follows: type II string theory on certain Calabi-Yau manifolds (``generalized conifolds'') is known to reduce to $\N=1$ Yang-Mills theories in a limit that decouples gravity; at low energies these geometrical spaces undergo a ``geometric transition'', where a cycle in the geometry shrinks to zero size and is replaced by a different cycle of finite size.  This is a geometrical analogue to confinement of the Yang-Mills theory at low energies.  If we instead consider B-type topological strings on these spaces, the topological string amplitudes reproduce the F-terms (superpotential) of the corresponding gauge theory.  Therefore, after the geometric transition they should give us the gauge theory effective superpotential.  However, the path integral of the topological B-model on these spaces reduces to a large $N$ matrix integral.  Following the chain of arguments, the effective superpotential of $\N=1$ Yang-Mills theories should reduce to a large $N$ matrix integral.  Thus, string theory provided an entirely unexpected computational tool for studying the effective superpotential of $\N=1$ gauge theories with matter.  

In practical terms, we can illustrate the technique as follows.  Suppose we start with a $\SU(N_c)$ gauge theory with $\N=1$ supersymmetry and a chiral superfield $\Phi$ in the adjoint representation, with a tree-level superpotential that contains a mass term and cubic self-interaction:

\begin{equation}
W = \int d^2 \theta~\left(\frac{m}{2} \Phi^{2} + \frac{g}{3} \Phi^3\right)
\end{equation}
String theory suggests that the effective superpotential of this theory $\Weff(S)$, written in terms of the gaugino bilinear $S$, receives contributions from two sources:  Veneziano-Yankielowicz terms arising from the strongly-coupled dynamics of the gauge field, and contributions from the matter field $\Phi$.  According to the conjecture, the only contributions of the matter field $\Phi$ to the effective superpotential come from the {\it planar} $\Phi$ diagrams of the theory (even at finite $N_c$) where we insert the external $S$ field once into each of the index loops of the $\Phi$ diagrams. Furthermore the effective superpotential has {\it no dependence} on the internal loop momenta of the diagrams!

The meaning of this result is that the superpotential for such theories is an essentially combinatorial object, depending only on the counting of ribbon diagrams with planar topology.  It has been known for a long time that these planar diagrams are counted by a zero-dimensional matrix integral \cite{Brezin:1978sv}, and we can often evaluate the free energy of this ``matrix model'' exactly.   The extension of this result to other gauge groups and matter content can be found in a long list of followup work to \cite{Dijkgraaf:2002fc}, including \cite{Ashok:2002bi,Janik:2002nz,Ita:2002kx,Aganagic:2003xq}; see \cite{Argurio:2003ym} for a review of field theoretical aspects.

We saw in the previous section that in non-supersymmetric field theories the need to integrate over loop momenta was a serious complication for extending the computation of the effective action to higher orders.  What is the field theory process that removes the contribution of loop integrals when supersymmetry is present?  As in non-supersymmetric theories, we can understand the field theory results in two ways: using anomalies \cite{Cachazo:2002ry} and by evaluating the path integral \cite{Dijkgraaf:2002xd}.  We will summarize the results of these papers, and refer to the original papers for the details.

The technique of using anomalous symmetries to solve for the effective superpotential has been extended to a large class of $\N=1$ theories \cite{Cachazo:2002ry, Seiberg:2002jq, Brandhuber:2003va}, where the relevant anomalies are of generalized Konishi type. This approach relies on the fact that the set of chiral primary fields -- those that can enter the effective superpotential -- are closed under addition and operator product, up to terms that vanish when evaluated in a supersymmetric vacuum; in other words the chiral primary fields generate a ring structure, the {\it chiral ring}.  Moreover, elements of the chiral ring are independent of position, so the chiral ring is a global structure.

Using the properties of the chiral ring, it was shown that the (anomalous) symmetries of the theory (particularly the generalized Konishi anomalies) restrict the possible superpotential contributions to the planar diagrams with insertions of $S$.  Then, the Ward identities associated to the generalized Konishi anomalies are shown to be equivalent to the loop equations of the matrix model, which are Dyson-Schwinger equations for the correlation functions, and which can be solved using matrix integral techniques to determine the effective superpotential exactly.

A complementary field theory approach \cite{Dijkgraaf:2002xd} used the background field method to study $\N=1$ gauge theories.  They showed that as a consequence of symmetries, it is again only the planar diagrams of the gauge theory that can contribute to the effective superpotential, and moreover supersymmetry implies that after the loop diagrams are summed in the Schwinger formalism, the loop momentum dependence in the diagram sum exactly cancels between bosonic and fermionic contributions.  Since there is no remaining dependence on loop momenta, the resulting effective superpotential reduces to the zero-dimensional matrix model calculation.  A key feature seen in this approach is that the individual gauge theory loop diagrams do depend on loop momenta, but after summing over all diagrams the momentum dependence exactly cancels.

There are several remarkable consequences of these results.  In many cases the associated matrix integral can be directly solved (corresponding to summing the Feynman diagram expansion to all orders).  However, in more complicated examples where the diagram series cannot easily be summed using known techniques, a perturbative expansion of the ribbon diagrams (up to some order in the number of index loops) gives a perturbative expansion of the effective superpotential $W(S)$, which upon extremization generates an expansion of the vacuum gluino condensate $\langle S \rangle \sim \langle \lambda \lambda \rangle$ as a sum of fractional instanton contributions.  As emphasized in \cite{Dijkgraaf:2002dh}, and as we have seen in other examples above, the perturbative loop expansion of the gauge theory in terms of an appropriate choice of composite operator yields non-perturbative information about the vacuum.

These results have been checked and extended in a large number of papers, and the deeper consequences for the quantum structure of gauge theories are still being explored. 

\section{Discussion}
\label{sec:disc}

As we have seen in several examples, the anomalous violation of scale invariance in massless quantum field theories gives rise to effective potentials of Coleman-Weinberg type \cite{Coleman:1973jx}.  The Callan-Symanzik equation, which imposes the violation of scale invariance by the trace anomaly, constrains the form of the possible quantum corrections to the classical potential, which are typically found to involve powers of $\log({\cal O}/\mu^{d_{\cal O}})$.  Due to these correction terms, the effective potential can have apparent non-trivial extrema.  These extrema are non-perturbative, in the sense that they lie at values of the classical field of the form

\begin{equation}
\langle {\cal O}\rangle = \mu^{d_{\cal O}} \exp(-\alpha/g^{2}) + \ldots
\end{equation}
which has strictly vanishing (or essentially singular) Taylor expansion around $g=0$. When other (anomalous, or non-anomalous) symmetries are present, they may further constrain the quantum corrections, in some cases being sufficient to determine the effective potential exactly. 

It is often the case that the 1-loop approximation breaks down in certain regions, because the first (and higher) quantum corrections become large, so these apparent non-trivial extrema may be removed or altered by higher loop corrections, but in some theories (such as the large $N$ Gross-Neveu model, and supersymmetric gauge theories), the 1-loop result is exact and the effective potential indeed has a non-trivial minimum.  In such cases, extremizing the effective potential gives exact non-perturbative information about the vacuum of the theory, such as the value of symmetry-breaking condensates.

{\bf Acknowledgements}

I am thankful to C.~Johnson, D.~Nemeschansky, G.~Sterman and N.~Warner for valuable discussions and critical reading of the manuscript, and to the members of the USC String Theory Journal Club for feedback.  This work was supported in part by DOE grant number DE-FG03-84ER-40168, and by a USC Final Year Dissertation Fellowship.

\bibliographystyle{utphys}
\bibliography{qft}

\end{document}